\documentclass[conference]{IEEEtran}
\IEEEoverridecommandlockouts
\usepackage{cite}
\usepackage{amsmath,amssymb,amsfonts}
\usepackage{graphicx}
\usepackage{textcomp}
\usepackage{xcolor}
\usepackage{setspace}
\usepackage{amsmath}
\usepackage{amsfonts}
\usepackage{bm}
\usepackage{mathtools}
\usepackage{hyperref}
\newtheorem{problem}{Problem}
\usepackage{algorithm}
\usepackage{algpseudocode}
\usepackage{booktabs}
\usepackage{makecell}
\usepackage{comment}
\usepackage{graphicx}
\usepackage{tikz}
\definecolor{goldenrod}{HTML}{FFDF42}

\def\BibTeX{{\rm B\kern-.05em{\sc i\kern-.025em b}\kern-.08em
    T\kern-.1667em\lower.7ex\hbox{E}\kern-.125emX}}

\begin{document}

\title{Optimal PMU Placement for Kalman Filtering of DAE Power System Models\\

\thanks{This work was supported as a part of NCCR Automation, a National Centre of Competence in Research, funded by the Swiss National Science Foundation (grant number 51NF40\_225155).}
}

\author{\IEEEauthorblockN{Milo\v{s} Katani{\'c}}
\IEEEauthorblockA{\textit{Power Systems Laboratory} \\
\textit{D-ITET, ETH}\\
Z{\"u}rich, Switzerland\\
mkatanic@ethz.ch}

\and
\IEEEauthorblockN{Yi Guo}
\IEEEauthorblockA{\textit{Power Systems Laboratory} \\
\textit{D-ITET, ETH}\\
Z{\"u}rich, Switzerland\\
yi.guo@empa.ch}
\and
\IEEEauthorblockN{John Lygeros}
\IEEEauthorblockA{\textit{Automatic Control Laboratory} \\
\textit{D-ITET, ETH}\\
Z{\"u}rich, Switzerland \\
jlygeros@ethz.ch}
\and
\IEEEauthorblockN{Gabriela Hug}
\IEEEauthorblockA{\textit{Power Systems Laboratory} \\
\textit{D-ITET, ETH}\\
Z{\"u}rich, Switzerland\\
hug@ethz.ch}

}

\maketitle

\begin{abstract}
Optimal sensor placement is essential for minimizing costs and ensuring accurate state estimation in power systems. This paper introduces a novel method for optimal sensor placement for dynamic state estimation of power systems modeled by differential-algebraic equations. The method identifies optimal sensor locations by minimizing the steady-state covariance matrix of the Kalman filter, thus minimizing the error of joint differential and algebraic state estimation. The problem is reformulated as a mixed-integer semidefinite program and effectively solved using off-the-shelf numerical solvers. Numerical results demonstrate the merits of the proposed approach by benchmarking its performance in phasor measurement unit placement in comparison to greedy algorithms.

\end{abstract}


\section{Introduction}

State estimation (SE) is an important monitoring tool for the daily operation of power systems \cite{abur}. A plethora of algorithms exist with widely distinct assumptions for static or dynamic SE. The performance of these algorithms heavily depends on sensor placement. The installation and maintenance costs of measurement sensors and the associated communication network can be substantial, making it economically impractical to deploy sensors at each power system node. Instead, careful consideration and strategic placement of sensors within the grid become imperative, not only to manage costs but also to ensure the effectiveness of the monitoring system. \par
Traditionally, optimal sensor placement for power systems has concentrated on minimising the error of static state estimation (SSE).
Techniques, such as genetic algorithms \cite{Milosevic2003}, topological schemes \cite{Baldwin1993}, and integer programming \cite{Xu2004}, have been employed for this purpose. These approaches typically disregard the dynamics and focus solely on algebraic network equations, optimizing the placement of conventional measurements (remote terminal units) and/or phasor measurement units (PMUs) with respect to specific metrics.\par
The increased availability of accurate high sampling rate PMUs enables the tracking of power system dynamics states, thus paving the way towards real-time, wide-area monitoring, protection, and control \cite{roles}. This evolution challenges the conventional static approach and enables dynamic state estimation (DSE) strategies for enhanced system understanding and responsiveness \cite{roles}. Several different metrics have been employed in the literature to quantify the observability of power systems for DSE. For instance, the Lie observability matrix is used in \cite{Lie_obs}, whereas \cite{c_dse2} and \cite{adapt_gram} employ the observability gramian. However, the referenced works are limited to placing PMUs only at generator nodes as they focus only on estimating the differential states of generators, neglecting the algebraic states. An alternative approach introduced in \cite{Sun2011} uses the Kalman filter steady-state estimation error covariance matrix as the PMU placement criterion. This method pre-selects potential PMU locations by solving the SSE problem and subsequently identifies the configuration with the smallest steady-state covariance matrix. The approach is only applicable, however, if DSE is performed in two steps: SSE and subsequent DSE for each generator. \par
Few studies have explored PMU placement for the joint estimation of both dynamic and algebraic states. Along this line, \cite{taha_PMU} approximates algebraic constraints by introducing dynamics to algebraic states using singular perturbation theory, hence transforming the original differential-algebraic equations (DAE) into approximated ordinary differential equations (ODE). This approach is limited to \mbox{index-1} DAE, which makes it inapplicable if the power system model is only partially known. In contrast, \cite{OPP_partial} considers partially unknown system models and addresses the PMU placement problem by minimizing the steady-state Kalman filter covariance matrix. The method relies on a simplified swing equation generator model and DC power flow, resulting in a simplified ODE power system model and simplified measurement functions. In addition, it employs a greedy algorithm without theoretical optimality guarantees, as the objective function is not submodular. \par
To the best of our knowledge, no existing algorithm provides optimality guarantees for determining the optimal PMU configuration that minimizes the steady-state covariance matrix in Kalman filtering for DAE-based power system models. This paper addresses this gap and makes the following key contributions:
\begin{itemize}
    \item We derive an exact reformulation for optimal sensor placement, aiming to minimize the trace of the steady-state covariance matrix obtained by the discrete-time algebraic Riccati equation for generalized DAE systems \cite{kalman_descriptor}. The problem is recast as a mixed-integer semi-definite program (MISDP), which can be effectively solved by off-the-shelf optimization packages. Our approach provides a novel, theoretically grounded, and computationally viable methodology for PMU placement in power systems.
    \item Our numerical results indicate that formulating the DSE using nodal injected currents as algebraic states offers a significant numerical advantage over using nodal voltages. While the use of nodal currents as states has been shown to accelerate the convergence of SSE \cite{currents_SSE}, to the best of our knowledge, it remains unexplored in the context of DAE-based DSE.
\end{itemize}
We apply two greedy algorithms from the literature to the considered problem, \textit{best-in} and \textit{worst-out}, and compare the results with the optimization-based approach. Thereby, we show the existence of realistic power system test cases where greedy algorithms exhibit a significant optimality gap. However, we also show that in many realistic test cases, greedy algorithms achieve near-optimal performance. 

\begin{section}{Methodology}
\label{sec:methodology}
\subsection{Power System Model}
The power system is modeled by the following set of differential-algebraic equations \cite{milano2010power}:
\begin{align}
    \dot{\bm{{{y}}}} &= \Tilde{\bm{f}}(\bm{{{y}}}, \bm{v}) + \Tilde{\bm{\xi}}_d,
    \label{eq:differential}\\
    \bm{0} &= \Tilde{\bm{g}}(\bm{{{y}}}, \bm{v}) + \Tilde{\bm{\xi}}_a,
    \label{eq:algebraic}
\end{align}
where $\bm{{{y}}}\in \mathbb{R}^{n_d}$ and $\bm{v} \in \mathbb{R}^{n_a}$ are, respectively, the vectors of differential states (internal generator states) and algebraic states (nodal voltages in phasor domain), $\Tilde{\bm{f}}: \mathbb{R}^{n_d + n_a} \rightarrow \mathbb{R}^{n_d} $ and $\Tilde{\bm{g}}: \mathbb{R}^{n_d + n_a} \rightarrow \mathbb{R}^{n_g} $ are nonlinear functions representing the time evolution and algebraic current or power flow equations, and $\Tilde{\bm{\xi}}_d \in \mathbb{R}^{n_d}$ and $\Tilde{\bm{\xi}}_a \in \mathbb{R}^{{n}_g}$ are continuous-time process noise vectors.\par 
Power systems are operated hierarchically by different stakeholders. While dynamic modeling may be feasible in regions with complete information (such as topology, parameters, and dynamic characteristics), it may be unavailable or unreliable in others, particularly in distribution networks. As a result, the model in \eqref{eq:differential}-\eqref{eq:algebraic} represents only the parts of the system with sufficient information. The remaining parts are treated as unknown, and their corresponding equations are omitted. Consequently, current or power balance equations at the interface nodes with these unknown regions are unavailable, i.e., $n_g \leq n_a$, leading to a system with more states than equations. For details on modeling partially known systems, we refer the reader to \cite{katanic2023} and \cite{OPP_partial}.
\par
We discretize the power system model using the implicit Euler method and linearize the equations around the nominal operating conditions \cite{katanic2023}, resulting in
\begin{align}
    \bm{E}\bm{x}_{k} &\approx \bm{A}\bm{x}_{k-1} + \bm{\Delta} + \bm{\xi}_k, 
    \label{eq:discrete}
\end{align}
where $\bm{x}_k \in \mathbb{R}^{n}$, with $n=n_d + n_a$, comprises both differential and algebraic states at time step $k$, $\bm{E} \in \mathbb{R}^{\tilde{n} \times n}$ and $\bm{A}\in \mathbb{R}^{\tilde{n} \times n}$ are the linearized system dynamic matrices, with $\tilde{n}=n_d+n_g$, $\bm{\Delta} \in \mathbb{R}^{\tilde{n}}$ is the linearization offset, and $\bm{\xi}_k\in \mathbb{R}^{\tilde{n}}$ is a process noise vector that we assume to be Gaussian zero mean with covariance $\bm{Q} \in \mathbb{S}^{\tilde{n}}_{++}$ and independent for different $k$. A measurement function $i$ is linear when phasor measurements are expressed in rectangular coordinates, where measured values are represented by the real and imaginary parts of the complex phasor \cite{linear_PMU}, and is given by
$    \bm{z}^i_{k} = \bm{C}_i \bm{x}_{k} + \bm{\nu}^i_{k}$, where $i = 1, \ldots, M$ denotes the sensor number, $\bm{z}^i_k \in \mathbb{R}^2$ is the measurement taken by sensor $i$, $\bm{C}_i \in \mathbb{R}^{2 \times n}$ is the $i$-th measurement matrix, and $\bm{\nu}^i_k \in \mathbb{R}^{2}$ is a measurement noise vector that we assume to be Gaussian noise process with zero-mean, covariance $\bm{R}_i \in \mathbb{S}^{2}_{++}$, and independent for different $k$ and from process noise. In this work, we focus solely on PMU measurements, even though the algorithm can be extended to include other measurement functions.\par

\subsection{Sensor Placement Algorithm}
The overall goal is to choose optimal locations for sensors to jointly minimize the error of the differential and algebraic states. This is achieved by minimizing the steady-state covariance matrix of the respective Kalman filter. For details about the underlying theory of Kalman filtering on descriptor systems, the reader is referred to \cite{kalman_descriptor}.\par
We collect all binary selection variables for $M$ distinct PMU sensors in a vector
$\bm{\gamma} := \{ \gamma_1, \gamma_2 ,\dots, \gamma_M\}  \in \{ 0, 1 \}^M $, where $1$ means the corresponding measurement is selected, and $0$ not selected. By defining the \textit{assimilated sensing precision matrix} $\bm{S} \in \mathbb{S}^{n}_+$ \cite{Yang2004} as:
\begin{equation}
    \bm{S}(\bm{\gamma}) = \sum_{i=1}^M\gamma_i\bm{C}^T_i\bm{R}_i^{-1}\bm{C}_i,
\end{equation}
the recursion of the Kalman filter covariance matrix $\bm{P}_k \in \mathbb{S}_{++}^{n}$ for generalized DAE models, can be expressed as \cite{ishihara}:
\begin{equation}
     \bm{P}_{k}(\bm{\gamma}) = \left[\bm{E}^T(\bm{Q} + \bm{A}\bm{P}_{k-1}(\bm{\gamma})\bm{A}^T)^{-1}\bm{E} + \bm{S(\gamma)}\right]^{-1}.
     \label{covariance_ekf}
\end{equation}
Under the assumptions specified in \cite{kalman_descriptor}, recursion (\ref{covariance_ekf}) converges to a unique positive definite matrix  $  \bm{P}_{\infty}(\bm{\gamma}) = \lim_{k \rightarrow \infty} \bm{P}_{k}(\bm{\gamma})$ known as the a posteriory steady state covariance matrix of the Kalman recursion.\par 

Let us define the operator that propagates the estimation covariance matrix one step into the future $\bm{g}: \mathbb{S}_{++}^n \times \{0,1\}^M \rightarrow \mathbb{S}_{++}^n$ as
$
\bm{g}(\bm{P}; \bm{\gamma}) = \left[\bm{E}^T(\bm{Q} + \bm{A}\bm{P}\bm{A}^T)^{-1}\bm{E} +\bm{S(\gamma)}\right]^{-1}.
$
The optimal PMU placement problem can now be formulated as:
\begin{problem}
\label{problem1}
\begin{subequations}
\begin{align}
   \min_{\bm{P},\bm{\gamma}} \quad &\textrm{Tr}(\bm{P})\\
    \textrm{s.t.} \quad &\bm{P} = \bm{g}(\bm{P}; \bm{\gamma}), \label{constr_lmi}\\
    &\bm{c}^T\bm{\gamma} \leq b,\\
    & \bm{P}\succ 0,\\
    &\bm{\gamma} \in \{ 0, 1 \}^M. \label{constr_int}
\end{align}
\end{subequations}
\end{problem}
Problem \ref{problem1} for linear ODE systems is known in the literature as the Kalman filter sensor selection problem (KFSS) \cite{Ye2018}. Here, $b \in \mathbb{R}_{+}$ represents the prespecified budget and $\bm{c} \in \mathbb{R}_{+}^{M}$ collects the individual measurement costs. In other words, we are looking for a positive-definite matrix with a minimal trace that satisfies the steady-state equation of the Kalman recursion and uses only PMU measurements with a cumulative cost smaller than the prespecified budget. \par 
Problem 1 is not solvable efficiently using any existing numerical algorithm because the feasible domain given by constraints (\ref{constr_lmi}) and (\ref{constr_int}) is not convex. Constraint $\bm{P} = \bm{g}(\bm{P}; \bm{\gamma})$ can be substituted by $\bm{P} \succeq \bm{g}(\bm{P}; \bm{\gamma})$ as at the optimal point, it will be satisfied with equality \cite{Yang2004}. From the Schur complement theorem, $\bm{P} \succeq \bm{g}(\bm{P}; \bm{\gamma})$ can be rewritten as
\begin{align}
    \begin{bmatrix}
        \bm{P} & \bm{I}\\
        \bm{I} & \bm{E}^T(\bm{Q} + \bm{A}\bm{P}\bm{A}^T)^{-1}\bm{E} + \bm{S(\gamma)}
    \end{bmatrix}\succeq 0.
\end{align}
Applying the matrix inversion lemma to $(\bm{Q} + \bm{A}\bm{P}\bm{A}^T)^{-1}$ yields
\begin{align}
    \left[\begin{array}{cc}
        \bm{P} & \bm{I} \\
        \bm{I} & \bm{M}
    \end{array}\right] \succeq 0,
    \label{eq:M}
\end{align}
where $\bm{M} = \bm{E}^T(\bm{Q}^{-1}-\bm{Q}^{-1}\bm{A}(\bm{P}^{-1}+\bm{A}^T\bm{Q}^{-1} \bm{A})^{-1} \bm{A}^T\bm{Q}^{-1})\bm{E}  + \bm{S(\gamma)}$. Reapplying the Schur complement theorem, this time to matrix $\bm{M}$ in \eqref{eq:M}, yields
\begin{equation}\bm{\Theta}=
    \begin{bmatrix} 
    \bm{P}^{-1} + \bm{A}^T\bm{Q}^{-1}\bm{A}& \bm{0} & \bm{A}^T\bm{Q}^{-1}\bm{E}\\
    \bm{0} & \bm{P} & \bm{I}\\
    \bm{E}^T\bm{Q}^{-1}\bm{A} & \bm{I}& \bm{E}^T\bm{Q}^{-1}\bm{E} + \bm{S(\gamma)}
    \end{bmatrix} \succeq   0,
\end{equation}
that is equivalent \cite{Sinopoli2004} to 
\begin{equation}
    \bm{\Lambda}=\begin{bmatrix}
        \bm{I} & \bm{0} & \bm{0}\\
        \bm{0} & \bm{P}^{-1} & \bm{0}\\
        \bm{0} & \bm{0} & \bm{I}
    \end{bmatrix}\bm{\Theta} \begin{bmatrix}
        \bm{I} & \bm{0} & \bm{0}\\
        \bm{0} & \bm{P}^{-1} & \bm{0}\\
        \bm{0} & \bm{0} & \bm{I}
    \end{bmatrix} \succeq 0.
\end{equation}
Introducing a change of variables $\bm{X}=\bm{P}^{-1}$, which can be enforced at the optimal point by an additional constraint  $\begin{bmatrix}
        \bm{P} & \bm{I}\\
        \bm{I} & \bm{X}
    \end{bmatrix}\succeq 0,$ 
we obtain an exact reformulation of \mbox{Problem \ref{problem1}}:
\begin{problem}
\begin{subequations}
\begin{align}
    \min_{\bm{P}, \bm{X}, \bm{\gamma}} & \textrm{Tr}(\bm{P}) \\
    \textrm{s.t.} \; \; \;& \! \!  \!\begin{bmatrix}
        \bm{P} & \bm{I} \\
        \bm{I} & \bm{X}
    \end{bmatrix} \succeq 0, \\
    & \! \!  \!\begin{bmatrix}
        \bm{X} + \bm{A}^T\bm{Q}^{-1}\bm{A}  & \bm{0} & \bm{A}^T\bm{Q}^{-1}\bm{E} \\
        \bm{0} & \bm{X} & \bm{X} \\
        \bm{E}^T\bm{Q}^{-1}\bm{A} & \bm{X} & \bm{E}^T\bm{Q}^{-1}\bm{E} + \bm{S(\gamma)}
    \end{bmatrix} \succeq 0, \\
    & \bm{P} \succ 0, \\
    & \bm{c}^T\bm{\gamma} \leq b, \\
    & \bm{\gamma} \in \{0, 1\}^M \label{eq:opt_bin}.
\end{align}
\end{subequations}
\end{problem}

Strictly speaking, this problem is still non-convex due to binary constraint (\ref{eq:opt_bin}), but relaxing it to the intervals $\bm{\gamma} \in [ 0, 1 ]^M$ transforms it into a convex problem. Hence, branch-and-bound algorithms, in combination with an SDP solver, can be employed to obtain a solution with optimality guarantees. The branch-and-bound algorithm systematically explores subsets of the solution space by dividing it into smaller subproblems (branching) and discarding regions that cannot contain the optimal solution (bounding). For each subproblem, the binary constraint is replaced by its relaxed form, allowing the SDP solver to solve the convex approximation. If the bound for a given subproblem is worse than the current best solution, the corresponding branch is pruned from further consideration. By iterating over the branching and bounding steps and progressively refining the bounds, the branch-and-bound algorithm converges to an optimal or near-optimal solution of the original problem.\par
The chosen PMU measurement configuration enables the reconstruction of all differential and algebraic states of the system with minimal steady-state error. As Kalman filtering, per design, minimizes the trace of the covariance matrix \cite{simon2006optimal}, the presented algorithm identifies among all possible sensor configurations (satisfying budget constraints) the one that minimizes the same metric. This selection thus represents the optimal sensor configuration for the optimal recursive filter. However, we acknowledge that the optimal sensor placement determined under one operating point may not remain optimal for different operating points, leaving room for future investigation.\par

\subsection{Alternative PMU Placement Methodologies}
A common alternative for sensor placement is the greedy approach, which iteratively adds sensors based on their contribution to reducing state estimation error. Greedy algorithms are widely used due to their simplicity and computational efficiency. Moreover, when the objective function is submodular, they guarantee a solution whose cost is within a factor of $1-e^{-1} \sim  0.63$ of the global optimum \cite{krause2014}. Submodularity can be defined as diminishing marginal return property, where adding an item to a smaller set provides more benefit than adding it to a larger set. Though some sensor and actuator placement problems that involve gramians are known to be submodular \cite{john_gramian}, it is shown in \cite{Zhang2017} that neither trace, determinant, nor inverse of the smallest eigenvalue of the steady state Kalman filter covariance matrix, are a submodular function in the selected sensor set for linear ODE systems. Consequently, greedy algorithms have no theoretical support in general; indeed, examples can be constructed where they perform arbitrarily poorly \cite{Ye2018}. Despite of this, they often perform very well in practice. \par
We consider here two variants of greedy algorithms: \textit{best-in} and \textit{worst-out} \cite{Ali2021}. The \textit{best-in} algorithm is initialized with the empty measurement set $\mathcal{T} = \emptyset$ and selects measurements by their contribution to the trace reduction of the steady state covariance matrix (see Algorithm~\ref{alg:greedy-in}), whereas \textit{worst-out} is initialized with the full measurement set $\mathcal{T} = \mathcal{P}$ and removes measurements one by one (see Algorithm~\ref{alg:greedy-out}).
\begin{algorithm}[!ht]

\caption{Greedy algorithm \textit{best-in}}
\begin{algorithmic}[1]
\State{Initialization: $l=1, \quad \mathcal{T} = \emptyset$ }
\While{$l \leq b$}
\For{ $i \in \mathcal{P} \cap \bar{\mathcal{T}}$}
\State{Calculate Tr$(\bm{P}_\infty(\mathcal{T}\cup \{i\}))$}
\EndFor
\State{$j = \arg \min_i \textrm{Tr} (\bm{P}_\infty(\mathcal{T}\cup \{i\}))$}
\State{$l \gets l+1,\quad \mathcal{T} \gets \mathcal{T} \cup \{j \}$}
\EndWhile
\end{algorithmic}
\label{alg:greedy-in}
\end{algorithm}

\begin{algorithm}[!ht]

\caption{Greedy algorithm \textit{worst-out}}
\begin{algorithmic}[1]
\State{Initialization: $l=M, \quad \mathcal{T} = \mathcal{P}$ }
\While{$l > b$}
\For{$i \in {\mathcal{T}}$}
\State{Calculate Tr($\bm{P}_\infty(\mathcal{T} \backslash \{i\}))$}
\EndFor
\State{$j = \arg \min_i \textrm{Tr} (\bm{P}_\infty(\mathcal{T} \backslash \{i\}))$}
\State{$l \gets l-1,\quad \mathcal{T} \gets \mathcal{T} \backslash \{j \}$}
\EndWhile
\end{algorithmic}
\label{alg:greedy-out}
\end{algorithm}

Evidently, \textit{best-in} has a computational advantage if the budget allows only a few sensors. On the other hand, \textit{worst-out} is better suited for incomplete power system models. In such cases, one PMU is not enough to achieve the convergence of the Riccati equation (as detailed in \cite{katanic2023}). Consequently, an initial measurement set must be chosen, which is not a trivial problem and cannot be addressed by the \textit{best-in} algorithm alone. \par

\end{section}
\section{Numerical Results}
\label{sec:numerical}
We validate the proposed algorithm on two test cases. In both cases, the generators are modeled by the fourth-order Anderson-Fouad's two-axis model, where the internal states consist of rotor angle, rotor speed, voltage behind transient reactance on the d-axis, and voltage behind transient reactance on the q-axis. The network is modeled by algebraic current balance equations \cite{milano2010power}. We assume that the inputs to synchronous generators, turbine power and exciter voltage are known. Future work will aim to remove this assumption and examine the placement under unknown inputs. All PMU measurements are assumed to be of unitary cost and are assumed to measure both the real and imaginary parts of the corresponding complex variable.\par
All simulations are implemented on an Intel i7-10510U CPU @ 1.80GHz with 16GB RAM. We employ the YALMIP optimization package \cite{Lofberg2004} and its branch-and-bound algorithm, and MOSEK \cite{mosek} to solve the SDPs. 
\subsection{3-bus system test case}
\label{subsec:small}
\begin{figure}
    \centering
    \includegraphics[width=0.42\columnwidth]{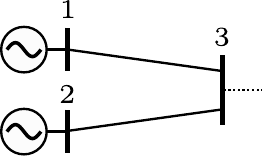}
    \caption{Single line diagram of the 3-bus system.}
    \label{fig:3bus}
\end{figure}

For illustrative purposes, we first analyze a small part of the power system as shown in Fig.~\ref{fig:3bus}. 
It comprises two synchronous generators connected to nodes 1 and 2. Node 3 can represent a distribution grid or an interconnection point to a larger power system, and it is assumed that the estimator has no information about the components ``behind'' node 3. Therefore, the available power system model is incomplete, and matrices $\bm{A}$ and $\bm{E}$ are rectangular (wide). For this analysis, the set of potential PMU measurements (comprising both the real and imaginary part of the corresponding complex value) is $\mathcal{P} = \{V_1, V_2, I_{2-3}\}$, where $V$ represents a voltage measurement, $I$ a current measurement, and subscripts denote the node or the branch. \par
\begin{figure}
    \centering
    \includegraphics[]{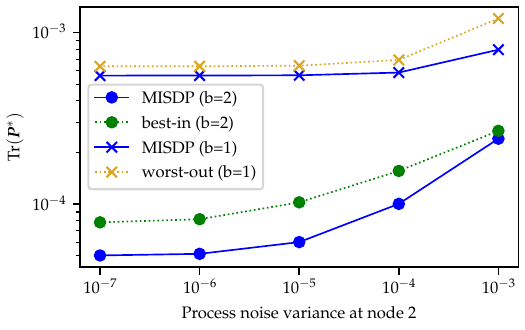}
    \caption{Steady state covariance matrix of the Kalman filtering for measurements selected by different algorithms evaluated under varying process noise at node 2.}
    \label{fig:toy_system}
\end{figure}

The results for the varying budget and varying process noise of algebraic current balance equation at node $2$ using the three different algorithms presented in Section~\ref{sec:methodology} are shown in Fig.~\ref{fig:toy_system}.
Note that the lower the trace of the Kalman filter covariance matrix, the more accurate the state estimation is. This noise magnitude can be interpreted as the uncertainty about other unmodeled injectors at node 2. \par
It is evident that both greedy \textit{best-in} and \textit{worst-out} achieve suboptimal performance. The optimal selection for the budget of one measurement is the voltage measurement at node 2. The greedy \textit{worst-out} algorithm removes this measurement in the first iteration. The optimal selection for the budget of two measurements is the voltage at node 1 and the current on the branch between nodes 2 and 3. In this case, the greedy best-in algorithm selects the voltage measurement at node 2 in the first iteration, which prevents it from identifying the optimal configuration after two iterations.\par
These results demonstrate that the proposed mixed-integer semi-definite program (MISDP) algorithm successfully identifies the optimal PMU configuration in scenarios where greedy algorithms yield only suboptimal solutions. In the following section, we evaluate the performance of both approaches on a larger test case, where we also discuss computational performance.

\subsection{11-bus system test case}
\begin{figure}
    \centering
    \includegraphics[width=0.7\columnwidth]{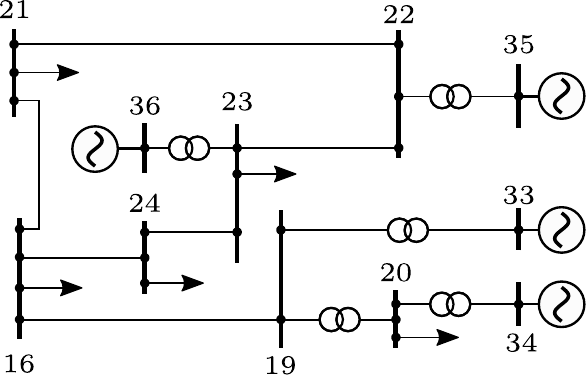}
    \caption{Single-line diagram of an 11-bus subsystem obtained by isolating a portion of the IEEE 39-bus test system.}
    \label{fig:11bus}
\end{figure}
The considered power system is shown in Fig.~\ref{fig:11bus} and comprises $11$ buses.\par

\subsubsection{Choice of Algebraic States}
Before addressing the optimal measurement location, we focus on the choice of algebraic states. Earlier works for SSE have shown that using injected currents in the phasor domain $\bm{i} \in \mathbb{R}^{n_a}$ as states can lead to improved convergence properties of the estimator \cite{currents_SSE}. Assuming that the system admittance matrix $\bm{Y} \in \mathbb{R}^{n_a \times n_a}$ is non-singular (see sufficient conditions in \cite{Low2022}), this approach corresponds to a simple physics-based linear coordinate transformation
$    \bm{i} = \bm{Y} \bm{v}$.\par
\begin{figure}
    \centering
    \includegraphics[width=\columnwidth]{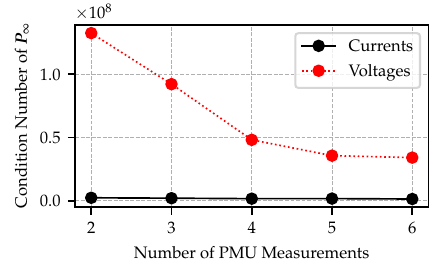}
    \caption{Average Kalman filter steady state covariance matrix for 20 random PMU measurement configurations with injected currents and injected voltages being the algebraic states.}
    \label{fig:cur_vs_volt}
\end{figure}

Fig.~\ref{fig:cur_vs_volt} shows the condition number of the steady state Kalman filter covariance matrix for the two choices of state variables for different numbers of measurements.  
For each number of measurements, 20 random PMU configurations are analyzed, and the average results are reported. The results clearly indicate that using currents as states not only improves the numerical properties of the estimator but also yields a more consistent condition number across varying numbers of measurements. This improvement suggests potential numerical advantages for both PMU placement and online recursive state estimation. Further investigations will explore the underlying causes and potential practical application of these findings.

\subsubsection{Optimal PMU locations}
For simplicity, we assume that all nodal voltages, nodal injected currents, and branch currents are candidate PMU measurements and can be selected independently. Therefore, the candidate measurements for this case are given by:
$ \mathcal{P} =\{$~$V_{16}$, $V_{23}$, $V_{21}$, $V_{24}$, $V_{20}$, $V_{22}$, $V_{19}$, $V_{33}$, $V_{34}$, $V_{35}$, $V_{36}$, $I_{16,19}$, $I_{20,19}$, $I_{16,24}$, $I_{16,21}$, $I_{22-21}$, $I_{22,23}$, $I_{34,20}$, $I_{33,19}$, $I_{23,24}$, $I_{23,36}$, $I_{22,35}$, $I_{19,16}$, $I_{19-20}$, $I_{24,16}$, $I_{21-16}$, $I_{21,22}$, $I_{23,22}$, $I_{20,34}$, $I_{19,33}$, $I_{24-23}$, $I_{36,23}$, $I_{35-22}$, $I_{20}$, $I_{23}$, $I_{24}$, $I_{16}$, $I_{21}$$\}$.\par
The results for optimal PMU placement under various budget constraints are summarized in Table~\ref{tab:sdp}, Table~\ref{tab:best-in}, and Table~\ref{tab:worst-out}. The findings indicate that all three algorithms produce similar measurement configurations. As expected, MISDP achieves the best performance, however, both greedy algorithms also demonstrate satisfactory results. Greedy \textit{best-in} selects a suboptimal solution for $b={3,4,5,6}$, whereas \textit{worst-out} selects a suboptimal choice for $b={2,3,4}$. Similar trends were observed across different process and measurement noise levels as well as varying operating points. These results suggest that greedy approaches provide a good trade-off between simplicity and performance, making them a viable and reasonable option in most realistic scenarios, however without theoretical guarantees.
\begin{table}
\centering
\caption{PMU placement and the associated cost optimized with \textbf{mixed-integer semi-definite programming (MISDP)} for varying budgets.}
\label{tab:sdp}
    \begin{tabular}{c|c l}
    \toprule
       b  & $\textrm{Tr}(\bm{P}^*_{gr,out})$& Selected PMU Measurements \\
       \midrule
       2&$3.380 \cdot 10^{-4}$&$V_{36}, I_{22-23}$\\
       3&$2.758 \cdot 10^{-4}$&$V_{36}, I_{23-36}, I_{22-35}$\\
       4&$2.565 \cdot 10^{-4}$&$V_{36}, I_{23-36} ,I_{22-35}, I_{33-19}$\\
       5&$2.493 \cdot 10^{-4}$&$V_{33}, I_{23-36}, I_{22-35}, I_{33-19}, I_{34-20},$\\
       6&$2.472 \cdot 10^{-4}$& $V_{33}, I_{23-36}, I_{22-35}, I_{33-19}, I_{34-20}, I_{16-19}$\\
       \bottomrule    
    \end{tabular}
\end{table}
\begin{table}
\centering
\caption{PMU placement and the associated cost optimized with \textbf{greedy best-in} for varying budgets.}
\label{tab:best-in}
\begin{tabular}{c|c l}
    \toprule
       b  & $\textrm{Tr}(\bm{P}^*_{MISDP})$& Selected PMU Measurements \\
       \midrule
       2&$3.380 \cdot 10^{-4}$&$V_{36}, I_{22-23}$\\
       3&$2.848 \cdot 10^{-4}$&$V_{36}, I_{22-23}, I_{16-19}$\\
       4&$2.661 \cdot 10^{-4}$&$V_{36}, I_{22-23}, I_{16-19}, I_{34-20}$\\
       5&$2.580 \cdot 10^{-4}$&$V_{36}, I_{22-23}, I_{16-19}, I_{34-20}, I_{22-35}$\\
       6&$2.504 \cdot 10^{-4}$& $V_{36}, I_{22-23}, I_{16-19}, I_{34-20}, I_{22-35}, I_{23-36}$\\
       \bottomrule     
    \end{tabular}
    
\end{table}
\begin{table}
\centering
\caption{PMU placement and the associated costs optimized with \textbf{greedy worst-out} for varying budgets.}
\label{tab:worst-out}
    \begin{tabular}{c|c l}
    \toprule
       b  & $\textrm{Tr}(\bm{P}^*_{gr,in})$& Selected PMU Measurements \\
       \midrule
       2&$3.693 \cdot 10^{-4}$&$ V_{33}, I_{23-36}$\\
       3&$2.779 \cdot 10^{-4}$&$V_{33}, I_{23-36}, I_{22-35}$\\
       4&$2.568 \cdot 10^{-4}$&$V_{33}, I_{23-36}, I_{22-35}, I_{33-19}$\\
       5&$2.493 \cdot 10^{-4}$&$V_{33}, I_{23-36}, I_{22-35}, I_{33-19}, I_{34-20}$\\
       6&$2.472 \cdot 10^{-4}$& $V_{33}, I_{23-36}, I_{22-35}, I_{33-19}, I_{34-20}, I_{16-19}$\\
       \bottomrule     
    \end{tabular}
\end{table}
We observe that all algorithms select one voltage measurement and the remaining branch current measurements, suggesting that current measurements may be more informative for the state reconstruction. This conclusion should be verified when the inputs to synchronous generators are kept unknown. The algorithm can be readily extended to incorporate more complex measurement functions, where combinations of multiple measurements can be installed at reduced costs compared to the sum of individual costs, reflecting economies of scale in installation and maintenance.\par
Now, we present the computational times observed under varying PMU budget scenarios. The results are summarized in Table~\ref{tab:comp_results} for all three algorithms. While the PMU placement technique proposed in this paper is designed as an offline approach—rendering computational time a secondary consideration—the execution times are important for scalability reasons. The MISDP running times for budgets of $b=5$ and $b=6$ PMUs are significantly shorter compared to other budgets, even outperforming the greedy \textit{worst-out} approach. This fact is likely due to an effective branching strategy. Overall, the greedy \textit{best-in} method demonstrates the fastest performance, particularly when the budget permits only a small number of sensors. To enhance scalability for the application to large power networks, future work will focus on optimizing the method’s computational efficiency.

\begin{table}[h!]
\centering
\caption{Computational performance of the presented algorithms for different PMU budgets.}
\label{tab:comp_results}
\begin{tabular}{l|c|c|c|c|c}
    \toprule
    Algorithm & \multicolumn{5}{c}{Time [s] for PMU Budget (b)} \\ 
    \cmidrule(lr){2-6}
    & 2 & 3& 4 & 5 & 6 \\
    \midrule
    MISDP      & 366 & 728 & 277 &  59 &  61 \\
    best-in   & 10 & 16 & 19 &  23 &  27 \\
    worst-out  & 89 & 82 & 79 &  71 &  77 \\
    
    \bottomrule
\end{tabular}
\end{table}

\section{Conclusions and Future Work}
\label{sec:conclusion}
This paper presents a novel method for optimal sensor placement for Kalman filtering of DAE models. The proposed method, formulated as a MISDP, ensures optimality and moderate computational times using off-the-shelf solvers. Numerical results confirm the method's applicability to PMU placement in realistic, small-size power system models. Additionally, we benchmark the performance of greedy algorithms against the proposed approach, showing that they achieve near-optimal performance and represent a viable choice for PMU placement but lack theoretical guarantees. \par
Future work should remove the assumption of known synchronous generators' inputs and perform the PMU placement with unknown generator inputs. Furthermore, the proposed method could benefit from numerical and computational enhancements. In summary, this work not only contributes a novel method for PMU placement but also opens up promising directions for advancing the understanding and application of joint dynamic and algebraic state estimation in power systems.

\bibliography{main.bib}

\begin{thebibliography}{10}

\bibitem{abur}
A.~Abur and A.~Exp{\'o}sito, {\em Power System State Estimation: Theory and Implementation}.
\newblock Power Engineering (Willis), CRC Press, 2004.

\bibitem{Milosevic2003}
B.~Milošević and M.~Begovic, ``Nondominated sorting genetic algorithm for optimal phasor measurement placement,'' {\em IEEE Transactions on Power Systems}, vol.~18, pp.~69--75, 02 2003.

\bibitem{Baldwin1993}
T.~Baldwin, L.~Mili, M.~Boisen, and R.~Adapa, ``Power system observability with minimal phasor measurement placement,'' {\em IEEE Transactions on Power Systems}, vol.~8, no.~2, pp.~707--715, 1993.

\bibitem{Xu2004}
B.~Xu and A.~Abur, ``Observability analysis and measurement placement for systems with pmus,'' in {\em IEEE PES Power Systems Conference and Exposition, 2004.}, pp.~943--946 vol.2, 2004.

\bibitem{roles}
J.~Zhao, M.~Netto, Z.~Huang, S.~S. Yu, A.~Gómez-Expósito, S.~Wang, I.~Kamwa, S.~Akhlaghi, L.~Mili, V.~Terzija, A.~P.~S. Meliopoulos, B.~Pal, A.~K. Singh, A.~Abur, T.~Bi, and A.~Rouhani, ``Roles of dynamic state estimation in power system modeling, monitoring and operation,'' {\em IEEE Transactions on Power Systems}, vol.~36, no.~3, pp.~2462--2472, 2021.

\bibitem{Lie_obs}
G.~Wang, C.-C. Liu, N.~Bhatt, E.~Farantatos, and M.~Patel, ``Observability of nonlinear power system dynamics using synchrophasor data,'' {\em Inter. Trans. on Electrical Energy Systems}, vol.~26, no.~5, pp.~952--967, 2016.

\bibitem{c_dse2}
J.~Qi, K.~Sun, and W.~Kang, ``Optimal {PMU} placement for power system dynamic state estimation by using empirical observability gramian,'' {\em IEEE Trans. on Power Systems}, vol.~30, no.~4, pp.~2041--2054, 2015.

\bibitem{adapt_gram}
J.~Qi, K.~Sun, and W.~Kang, ``Adaptive optimal pmu placement based on empirical observability gramian,'' vol.~49, no.~18, pp.~482--487, 2016.
\newblock 10th IFAC Symposium on Nonlinear Control Systems.

\bibitem{Sun2011}
Y.~Sun, P.~Du, Z.~Huang, K.~Kalsi, R.~Diao, K.~K. Anderson, Y.~Li, and B.~Lee, ``Pmu placement for dynamic state tracking of power systems,'' in {\em 2011 North American Power Symposium}, pp.~1--7, 2011.

\bibitem{taha_PMU}
M.~H. Kazma and A.~F. Taha, ``Optimal placement of {PMU}s in power networks: Modularity meets a priori optimization,'' in {\em 2023 American Control Conference (ACC)}, pp.~4489--4494, 2023.

\bibitem{OPP_partial}
M.~A. Abooshahab, M.~Hovd, and G.~Valmorbida, ``{Optimal Sensor Placement for Partially Known Power System Dynamic Estimation},'' in {\em {IEEE PES ISGT Europe 2021}}, (Espoo, Finland), Oct. 2021.

\bibitem{kalman_descriptor}
R.~Nikoukhah, A.~Willsky, and B.~Levy, ``{K}alman filtering and {R}iccati equations for descriptor systems,'' {\em IEEE Transactions on Automatic Control}, vol.~37, no.~9, pp.~1325--1342, 1992.

\bibitem{currents_SSE}
B.~K. Poolla, G.~Cavraro, and A.~Bernstein, ``State estimation for distribution networks with asynchronous sensors using stochastic descent,'' in {\em IEEE Power \& Energy Society General Meeting}, pp.~1--5, 2022.

\bibitem{milano2010power}
F.~Milano, {\em Power System Modelling and Scripting}.
\newblock Power Systems, Springer Berlin Heidelberg, 2010.

\bibitem{katanic2023}
M.~Katanic, J.~Lygeros, and G.~Hug, ``Recursive dynamic state estimation for power systems with an incomplete nonlinear dae model,'' {\em IET Gen., Trans. \& Distribution}, vol.~18, no.~22, pp.~3657--3668, 2024.

\bibitem{linear_PMU}
M.~Göl and A.~Abur, ``{LAV} based robust state estimation for systems measured by {PMU}s,'' {\em IEEE Transactions on Smart Grid}, vol.~5, no.~4, pp.~1808--1814, 2014.

\bibitem{Yang2004}
C.~Yang, J.~Wu, X.~Ren, W.~Yang, H.~Shi, and L.~Shi, ``Deterministic sensor selection for centralized state estimation under limited communication resource,'' {\em IEEE Transactions on Signal Processing}, vol.~63, no.~9, pp.~2336--2348, 2015.

\bibitem{ishihara}
J.~Ishihara, J.~Campos, and M.~Terra, ``Optimal recursive estimation for discrete-time descriptor systems,'' in {\em Proceedings of the 2004 American Control Conference}, vol.~1, pp.~188--193 vol.1, 2004.

\bibitem{Ye2018}
L.~Ye, S.~Roy, and S.~Sundaram, ``On the complexity and approximability of optimal sensor selection for kalman filtering,'' in {\em 2018 Annual American Control Conference (ACC)}, pp.~5049--5054, 2018.

\bibitem{Sinopoli2004}
B.~Sinopoli, L.~Schenato, M.~Franceschetti, K.~Poolla, M.~Jordan, and S.~Sastry, ``Kalman filtering with intermittent observations,'' {\em IEEE Transactions on Automatic Control}, vol.~49, no.~9, pp.~1453--1464, 2004.

\bibitem{simon2006optimal}
D.~Simon, {\em Optimal State Estimation: Kalman, H Infinity, and Nonlinear Approaches}.
\newblock Wiley, 2006.

\bibitem{krause2014}
A.~Krause and D.~Golovin, ``Submodular function maximization.,'' {\em Tractability}, vol.~3, no.~71-104, p.~3, 2014.

\bibitem{john_gramian}
T.~H. Summers, F.~L. Cortesi, and J.~Lygeros, ``On submodularity and controllability in complex dynamical networks,'' {\em IEEE Transactions on Control of Network Systems}, vol.~3, no.~1, pp.~91--101, 2016.

\bibitem{Zhang2017}
H.~Zhang, R.~Ayoub, and S.~Sundaram, ``Sensor selection for kalman filtering of linear dynamical systems: Complexity, limitations and greedy algorithms,'' {\em Automatica}, vol.~78, pp.~202--210, 2017.

\bibitem{Ali2021}
M.~Ali~Abooshahab, M.~Hovd, and G.~Valmorbida, ``Optimal sensor placement for partially known power system dynamic estimation,'' in {\em Innovative Smart Grid Technologies Europe}, pp.~1--6, 2021.

\bibitem{Lofberg2004}
J.~L{\"{o}}fberg, ``Yalmip : A toolbox for modeling and optimization in matlab,'' in {\em In Proceedings of the CACSD Conference}, 2004.

\bibitem{mosek}
M.~ApS, {\em The MOSEK optimization toolbox for MATLAB manual. Version 10.0.}, 2022.

\bibitem{Low2022}
S.~H. Low, ``Lecture notes for power system analysis - a mathematical aproach,'' 2022.

\end{thebibliography}
\bibliographystyle{ieeetr}

\end{document}